\documentclass[aps,prl,twocolumn,superscriptaddress]{revtex4}
\usepackage{graphicx}

\usepackage{dcolumn}

\usepackage{bm}
\usepackage{color}
\usepackage[colorlinks]{hyperref}
\input{epsf}
\begin{document}

%\preprint{}

\title{Radiation-induced zero-resistance states with resolved Landau levels}

\author{R. G. Mani}
\affiliation{Harvard University, Gordon McKay Laboratory of
Applied Science, 9 Oxford Street, Cambridge, MA 02138, USA}

%\author{}
%\thanks{}
%\affiliation{}
\date{\today}
\begin{abstract}
The microwave-photoexcited high mobility GaAs/AlGaAs
two-dimensional electron system exhibits an
oscillatory-magnetoresistance with vanishing resistance in the
vicinity of magnetic fields $B = [4/(4j+1)] B_{f}$, where $B_{f} =
2\pi\textit{f}m^{*}/e$, m$^{*}$ is an the effective mass, e is the
charge, \textit{f} is the microwave frequency, and $j$ =1,2,3...
Here, we report transport with well-resolved Landau levels, and
some transmission characteristics. \vspace{3 mm}

\noindent {Journal Reference: Appl. Phys. Lett. \textbf{85}, 4962
(2004).}
\end{abstract}
%\pacs{73.21.-b,73.40.-c,73.43.-f; Journal Ref: Appl. Phys. Lett.
%\textbf{85}, 4962 (2004).}
%\pacs{}
%\keywords{}
\maketitle

The experimental study of quantized Hall effect (QHE) has shown
that a two-dimensional electron system (2DES) can exhibit
zero-resistance states in the vicinity of integral and mostly
odd-denominator fractional filling factors, at low temperatures,
$T$, and high magnetic fields, $B$. These (quantum Hall) zero -
resistance states are initially approached, at finite $T$,
following an activation law, which is understood as a
manifestation of a gap in the electronic spectrum.\cite{1}  Recent
observations of radiation-induced zero-resistance states in the
2DES are particularly interesting because they have shown that the
above mentioned characteristics, i.e., activated transport and
zero-resistance states, can also be obtained in a photo-excited
high mobility 2DES, without realizing at the same time a
QHE.\cite{2,3} In such a situation, one wonders whether the
observed characteristics could once again be indicative of a
spectral gap, as in the quantum Hall limit.\cite{1}

The zero-resistance states of interest here are induced by
microwave excitation of a high mobility GaAs/AlGaAs 2DES, at low
$T$, in a large filling factor limit.\cite{2,3,4} Experiments
indicate vanishing diagonal resistance, following an activation
law, about $B$ = (4/5)$B_{f}$ and $B$ = (4/9)$B_{f}$, where
$B_{f}$ = $2\pi f m^{*}/e$, $m^{*}$ is an effective mass, $e$ is
the electron charge, and $f$ is the radiation frequency.\cite{2}
In this report, we illustrate transport with resolved Landau
levels and transmission characteristics, as we refer the reader to
the literature for discussions of theory.\cite{5,6}

Experiments were performed, as indicated elsewhere,\cite{2} on
standard devices fabricated from GaAs/AlGaAs heterostructure
junctions with an electron mobility up to $1.5$ $\times$ $10^{7}$
cm$^{2}$/Vs.  Typically, a specimen was mounted inside a
waveguide, immersed in pumped liquid Helium, and irradiated with
electromagnetic (micro-) waves over the frequency range $27 \leq f
\leq 170$ GHz. Reported electrical measurements were carried out
using low frequency \textit{ac} lock-in techniques, as usual.

\begin{figure}[t]
%h=here, t=top, b=bottom, p=separate figure page
\begin{center}
\leavevmode \epsfxsize=3.25in
 \epsfbox {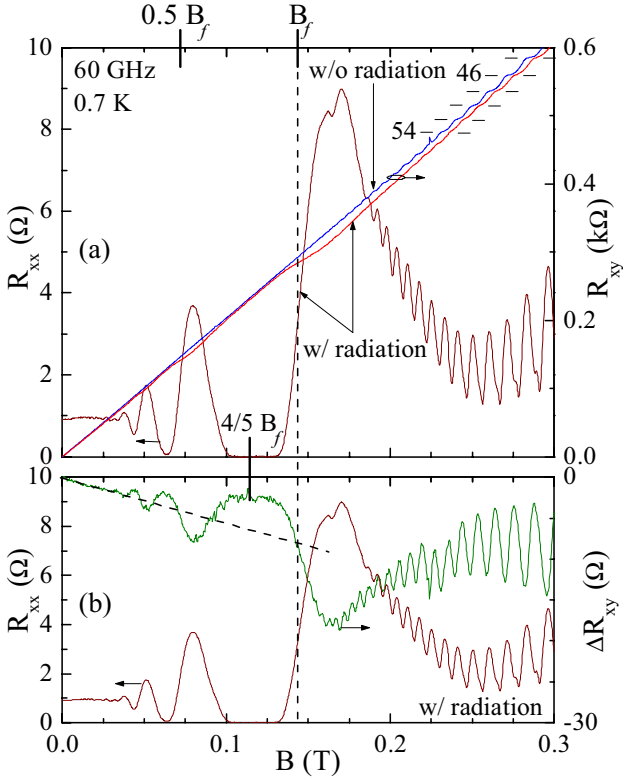}
\end{center}
\caption{(Color online) The figure exhibits the Hall resistance
$R_{xy}$ observed both with (w/) and without (w/o) microwave
radiation at $f$ = 60 GHz, along with $R_{xx}$ under radiation.
Here, $R_{xx}$ vanishes about $(4/5) B_{f}$, and this
zero-resistance state does not produce a plateau in $R_{xy}$. Note
the remarkable shift of QHE plateaus, i.e., plateaus with $R_{xy}
= h/i e^{2}$, for 46 $\leq$ $i$ $\leq$ 54, to higher B under the
influence of radiation. (b): The radiation induced correction to
the Hall resistance, $\Delta R_{xy}$, is shown along with
$R_{xx}$. Notably, the $\Delta R_{xy}$ oscillations correspond to
a decrease in the magnitude of $R_{xy}$, while the slope in
$\Delta R_{xy}$ (dotted line) indicates a change in the slope of
the $R_{xy}$ curve upon irradiation. The latter feature is
qualitatively consistent with the shift of the QHE plateaus to
higher $B$ under radiation, see (a).} \label{mani01fig}
\end{figure}

Figure 1 shows the low-$B$ transport under photoexcitation at 60
GHz. Fig. 1(a) indicates a wide $R_{xx}$ radiation-induced
zero-resistance state about (4/5)$B_{f}$, and a close approach to
vanishing resistance at the next lower-$B$ minimum, near
(4/9)$B_{f}$, which follow the series $B = (4/[4j+1])B_{f}$, with
$j$ = 1,2,3...\cite{2} A remarkable feature in these
zero-resistance states is the absence of concomitant plateau
formation in the Hall effect,\cite{2,3} as in typical quantum Hall
systems.\cite{1} The $R_{xy}$ data of Fig 1(a) show, for example,
that the Hall resistance increases linearly vs. the magnetic field
and coincides with the $R_{xy}$ that is observed without radiation
over the $B$ interval corresponding to the (4/5)$B_{f}$
zero-resistance state. Yet, a careful comparison of the
photoexcited (w/ radiation) Hall effect with the dark (w/o
radiation) Hall effect reveals some definite modifications in
$R_{xy}$ upon irradiation. For example, at the $R_{xx}$ maxima,
there appear to be reductions in the magnitude of the irradiated
$R_{xy}$. In addition, well above $B_{f}$, i.e., $B$ $\geq$ 0.2
Tesla, where $R_{xy}$ exhibits QHE plateaus, a given filling
factor QHE appears shifted to higher $B$ under the influence of
radiation. These effects could, however, be reversed simply by
switching off the microwave excitation.

\begin{figure}[t]
%h=here, t=top, b=bottom, p=separate figure page
\begin{center}
\leavevmode \epsfxsize=3.25in
 \epsfbox {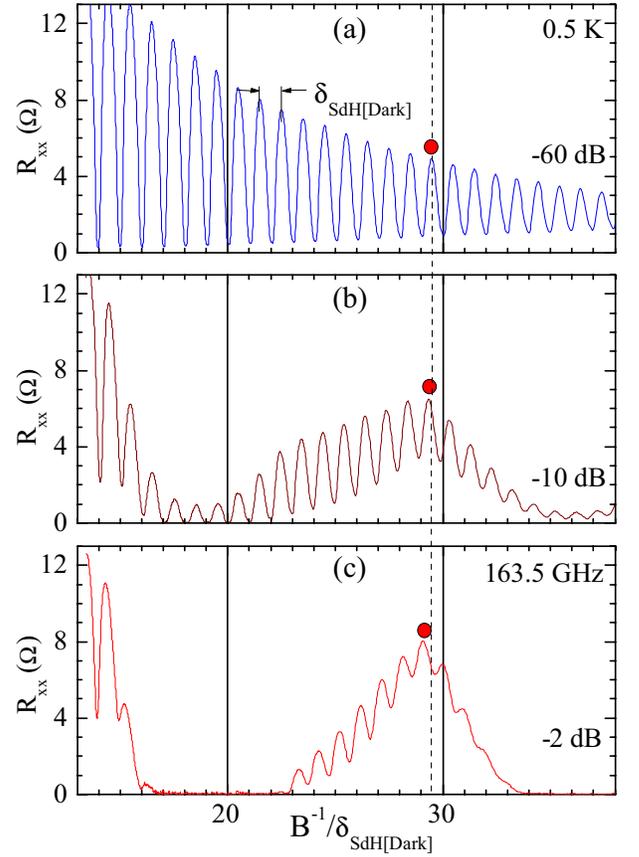}
\end{center}
\caption{(Color online) This figure shows the evolution of the
diagonal resistance $R_{xx}$ under radiation, in the regime where
large amplitude Shubnikov-de Haas (SdH) oscillations are
observable in $R_{xx}$. Top: This panel shows the SdH oscillations
in the absence of radiation (-60 dB). The abscissa has been
normalized by the period $\delta_{SdH [Dark]}$ of these SdH
oscillations. Center: Here, $R_{xx}$ under 163.5 GHz excitation
has been exhibited  with the radiation intensity attenuated to -10
dB. The radiation produces a strong modulation in the amplitude of
the SdH oscillations, which is a signature of the
radiation-induced oscillatory magnetoresistance. Bottom: $R_{xx}$
under photoexcitation, with the radiation intensity attenuated to
-2 dB. Note the radiation induced zero-resistance states about 17
$\leq$ $B^{-1}/\delta_{SdH [Dark]}$ $\leq$ 23 and 34 $<$
$B^{-1}/\delta_{SdH [Dark]}$, where the SdH oscillations also
vanish. As the colored disks mark a fixed filling factor, their
shift along the abscissa between (a) - (c) is interpreted as a
change in the cross sectional area of the Fermi surface under
radiation.} \label{mani03fig}
\end{figure}

\begin{figure}[t]
%h=here, t=top, b=bottom, p=separate figure page
\begin{center}
\leavevmode \epsfxsize=3.25in
 \epsfbox {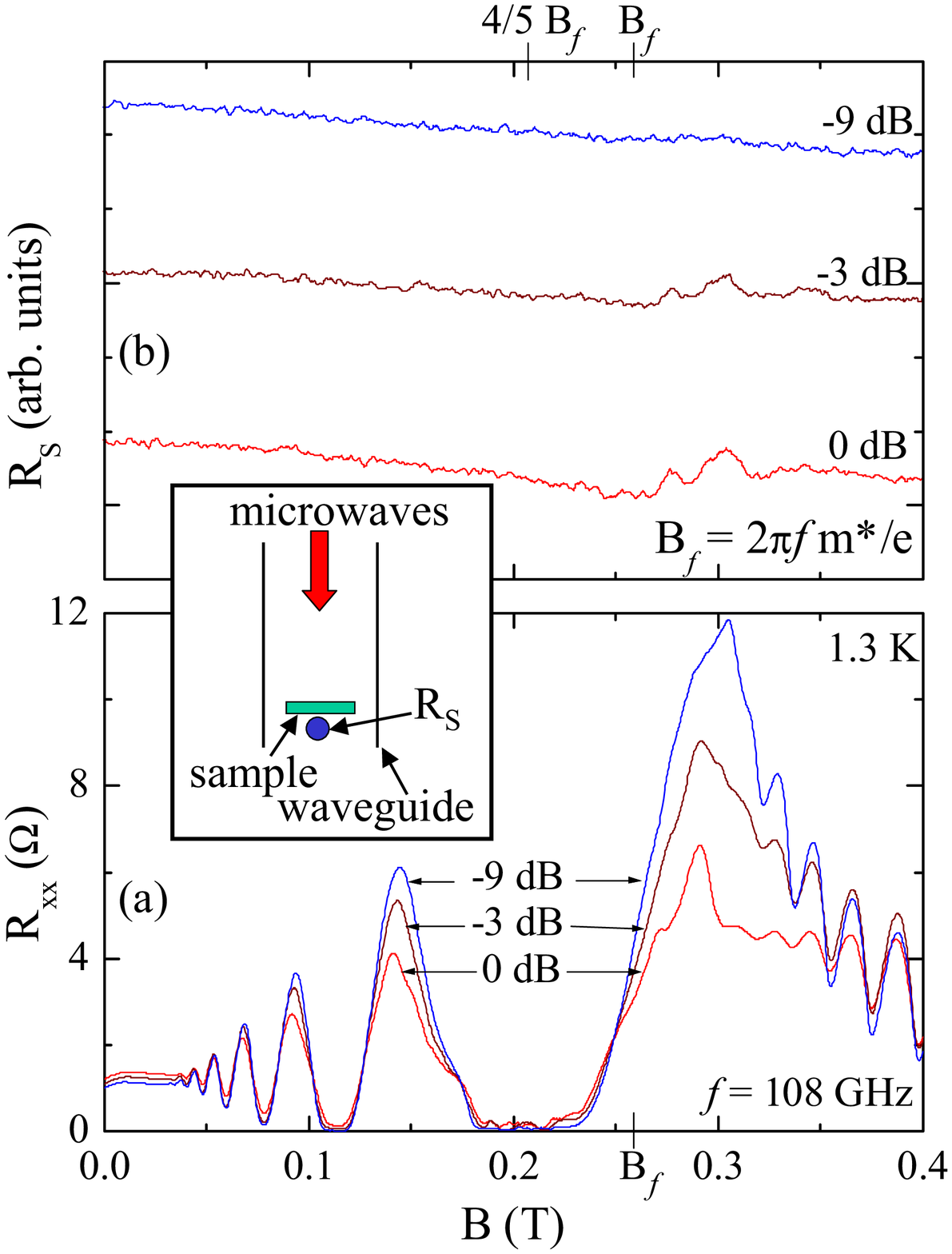}
\end{center}
 \caption{(Color online) This figure illustrates the transmission characteristics
of the 2DES under irradiation. Inset: A resistance sensor below
the sample serves as the radiation detector. (a) $R_{xx}$ of the
2DES vs. $B$.  The oscillation amplitude decreases with increasing
radiation intensity when the power attenuation factor exceeds
-9dB, signifying "breakdown". (b) The magnetoresistance of the
radiation sensor. This detector response suggests nonmonotonic
transmission above $B_{f}$.} \label{mani04fig}\end{figure}

As suggested by Fig. 1, radiation-induced zero-resistance states
are typically observed over a range of magnetic fields
corresponding to weak- or non-existent SdH oscillations (in the
dark) at easily accessible microwave frequencies and temperatures.
A question that we wish to address is whether radiation-induced
zero-resistance states can also occur over the range of $B$ where
the amplitude of SdH oscillations in the dark is relatively large,
say substantially greater than one-half of the \textit{background}
dc resistance in the absence of microwave excitation, at the same
B.

We show here results which indicate that, indeed, these
radiation-induced zero-resistance states can also occur in the $B$
limit, where giant Shubnikov-de Haas oscillations are observable
in the specimen. That is, where the Landau level spacing, $\hbar
\omega_{C}$, exceeds both the thermal energy, $k_{B}T$, and a
broadening parameter, $\Gamma$, defined from the transport
relaxation time i.e., $\Gamma << k_{B}T < \hbar \omega_{C}$, which
may be viewed as a quantum Hall threshold. This is a regime of
consequence because theory has sometimes identified observed
effects with the limit where $\Gamma << k_{B}T \approx \hbar
\omega_{C}$.\cite{6}

Fig. 2(a) shows that the dark specimen exhibits strong SdH
oscillations over the range of $B^{-1}$ given by 13 $\leq$
$B^{-1}/\delta_{SdH[Dark]}$ $\leq$ 38. This corresponds to filling
factors 26 $\leq$ $\nu$ $\leq$ 76 because a factor-of-two occurs
between $\nu$ and $B^{-1}/\delta_{SdH[Dark]}$, i.e., $\nu$ =
2$B^{-1}/\delta_{SdH[Dark]}$, when spin splitting is not resolved
in the SdH oscillations. Here, at the examined temperature, $T$ =
0.5 K, $hf$ = 0.676 meV easily exceeds $k_{B}T$ = 0.043 meV.
Estimates of the broadening parameter indicate that $\Gamma <<
k_{B}T$.\cite{2}

Photoexcitation of the specimen with $f$ = 163.5 GHz radiation
initially produces a modulation in the amplitude of the SdH
oscillations (see Fig. 2(b)), which is the signature of the
radiation-induced resistance oscillations in this separated Landau
level limit. One might plausibly explain this SdH modulation
feature, at least about the radiation induced resistance minima,
by suggesting that the radiation reduces the background diagonal
resistivity, and that in turn suppresses the amplitude of SdH
oscillations. Here, we imagine that the background resistance (and
resistivity) could be defined (and extracted) from the midpoints
of the SdH oscillations.

A further increase in the radiation intensity, see Fig. 2(c),
leads to zero-resistance states over broad $B^{-1}$ intervals. For
example, the (4/5)$B_{f}$ state occurs here about $B^{-1} /
\delta_{SdH[Dark]}$ = 20, and it looks similar to the effect that
is observed at lower $f$ (see Fig. 1). Note that at (4/5)$B_{f}$,
$\hbar \omega_{C}$ [= (4/5)$hf$] is  noticeably greater than
$k_{B}T$. A remarkable feature in these data of Fig. 2(c) is that
the SdH oscillations disappear as $R_{xx}$ $\longrightarrow$ 0
under the influence of radiation. It appears worth pointing out
that we have not observed any evidence of "chopping" of the SdH
minima on the approach to zero-resistance, as might be expected if
the amplitude of the SdH oscillations did not appropriately follow
the background $R_{xx}$, or if the SdH amplitude somehow stayed
constant as the background resistivity went to zero under
microwave excitation.

It is also worth considering the SdH behavior outside of the
domain of zero-resistance states. For example, on either side of
the $(4/5)B_{f}$ zero-resistance state (about
 $B^{-1}/\delta_{SdH[Dark]}$ = 20), on the adjacent $R_{xx}$
maxima, SdH oscillations continue to be observable, even as they
have disappeared on the zero-resistance states themselves. Note
that as the background resistance is enhanced with respect to the
dark value near, for example, $B^{-1}/\delta_{SdH[Dark]}$ = 29 in
Fig. 2(c), the amplitude of the SdH oscillations is not also
increased, and this suggests a break in the correlation between
the background resistance and the SdH amplitude at the maxima,
unlike the case with the minima. A noteworthy point here seems to
be that in Fig. 2 (b) and (c), the SdH oscillations seem not to
increase in amplitude under the influence of microwaves. The SdH
amplitude either stays the same or it is reduced under microwave
excitation. This might indicate a role for electron heating.
Parenthetically, there is also some evidence that the threshold
radiation intensity for realizing zero-resistance increases, as
one moves to higher $f$. Thus, heating could be more influential
at higher excitation frequencies. This feature is attributed here
to the point that, as the photon energy increases with $f$, more
power needs to be delivered to the specimen in order to maintain a
constant photon number per unit of time, which could be the
essential underlying parameter, at a higher $f$. A subtle feature
of interest in Fig. 2 is that the SdH extrema seem to shift to
lower $B^{-1}$ (or higher $B$) under microwave excitation, in
qualitative agreement with the behavior observed in the Hall
effect in Fig. 1(a).\cite{2}

The characteristic field scale for the radiation induced effect,
$B_{f}$,\cite{2} suggests a possible relation to cyclotron
resonance, which one might investigate through simultaneous
transmission and transport measurements in the same high mobility
specimen, see Fig. 3.  Here (see Fig. 3(inset)), a resistance
sensor placed immediately below the sample served to gauge the
relative transmitted power. Fig. 3(a) illustrates the specimen
$R_{xx}$ vs. $B$, while Fig. 3(b) exhibits the $B$-dependent
sensor resistance, $R_{S}$. In this GaAs/AlGaAs specimen, the
optimal radiation induced $R_{xx}$ response occurred in the
vicinity of -9 dB, see Fig. 3(a). That is, the amplitude of the
radiation-induced resistance oscillations increased monotonically
with increasing power up to -9 dB. A further increase of the
radiation intensity (dB $\rightarrow$ 0) produced a "breakdown" or
a reduction in the $R_{xx}$ peak height along with an
\textit{increase} in the resistance at the minima (see Fig. 3(a)).
The response of the transmission sensor, $R_{S}$, (see Fig. 3(b))
suggests structure at magnetic fields about- and mostly above-
$B_{f}$, which becomes more pronounced with increased excitation.
The feature correlates with a strong radiation-induced distortion
of the $R_{xx}$ peak that is centered about 0.3 Tesla. One might
interpret some of these features about $B_{f}$ as a signature of
cyclotron resonance, although further supplementary evidence
appears necessary to confirm this hypothesis. Remarkably, the
observed oscillations in $R_{xx}$ below $B_{f}$ appear
imperceptible in the sensor response (cf. Fig. 3(a) and Fig.
3(b)).

In summary, we have emphasized the possibility of realizing
radiation induced zero-resistance states, see Fig. 2, in a range
of $B$ where $R_{xx}$ exhibits giant SdH oscillations due to
separated Landau levels. This result indicates that
radiation-induced zero-resistance states occur even outside a
so-called quasi-classical limit. Transmission measurements also
indicate non-monotonic features in the transmitted signal about
and above $B_{f}$, some of which could be indicative of cyclotron
resonance.

The authors acknowledge discussions with K. von Klitzing, V.
Narayanamurti, J. H. Smet, and W. B. Johnson. The high mobility
material was kindly provided by V. Umansky. \vspace{0cm}

\end{document}